\documentclass[showpacs, aps, prb, unsortedaddress]{revtex4-1}

\usepackage{amssymb}
\usepackage{latexsym}
\usepackage{amsmath}
\usepackage{graphicx}

\begin{document}

\title{Causality and Passivity in Elastodynamics}

\date{\today}
\author{Ankit Srivastava}
\thanks{Corresponding Author}
\email{ankit.srivastava@iit.edu}

\affiliation{Department of Mechanical, Materials, and Aerospace Engineering
Illinois Institute of Technology, Chicago, IL, 60616
USA}

\begin{abstract}
What are the constraints placed on the constitutive tensors of elastodynamics by the requirements that the linear elastodynamic system under consideration be both causal (effects succeed causes) and passive (system doesn't produce energy)? The analogous question has been tackled in other areas but in the case of elastodynamics its treatment is complicated by the higher order tensorial nature of its constitutive relations. In this paper we clarify the effect of these constraints on highly general forms of the elastodynamic constitutive relations. We show that the satisfaction of passivity (and causality) directly requires that the hermitian parts of the transforms (Fourier and Laplace) of the time derivatives of the constitutive tensors be positive semi-definite. Additionally, the conditions require that the non-hermitian parts of the Fourier transforms of the constitutive tensors be positive semi-definite for positive values of frequency. When major symmetries are assumed these definiteness relations apply simply to the real and imaginary parts of the relevant tensors. For diagonal and one-dimensional problems, these positive semi-definiteness relationships reduce to simple inequality relations over the real and imaginary parts, as they should. Finally we extend the results to highly general constitutive relations which include the Willis inhomogeneous relations as a special case. 
\end{abstract}

\maketitle

\section{Introduction}
Various aspects of nature are modeled as cause-effect relationships between different physical processes. These physical processes are often functions of time and the relations between them, in some cases, can be more easily analyzed in the frequency or Laplace domains. A frequency dependent process is called \emph{dispersive} and can be studied by deriving the appropriate dispersion relations of the system. Physical systems often have an inherent assumption of causality wherein effects are assumed not to precede causes. If the physical system is also linear and time-invariant then certain sum/integral rules could be derived connecting the physical quantities involved \cite{titchmarsh1948introduction,toll1956causality}. For eg. the Kramers-Kronig (K-K) relationships \cite{kronig1926theory,kramers1927diffusion} are integral relationships which connect the real part of the electromagnetic index of refraction to its imaginary part, thus connecting dispersion and loss in the medium. Since their introduction, the K-K relationships have been used in the study of circuit theory \cite{zemanian1972realizability} and all forms of wave propagation \cite{jackson1962classical,nussenzveig1972causality,bode1940relations,feenberg1932scattering,ginzberg1955concerning,mangulis1964kramers,o1978general,booij1982generalization}. 

The K-K relations have recently attracted interest in the area of metamaterials where the goal is to create materials with exotic electromagnetic, acoustic, and/or elastodynamic properties. The essential ideas emerge from early theoretical works of Veselago \cite{veselago1968electrodynamics} and more recent experimental efforts by various research groups\cite{smith2000composite,shelby2001experimental,liu2000locally} (See \cite{srivastava2015elastic} for a review). The possibility of creating materials with unprecedented material properties has led to far-reaching postulations of their applications, most visibly, in the area of cloaking \cite{greenleaf2003anisotropic,greenleaf2003nonuniqueness,leonhardt2006notes,leonhardt2006optical,pendry2006controlling,norris2008acoustic,milton2006cloaking,norris2011elastic}. Since material properties can essentially be viewed as time-domain transfer functions which relate a cause to its effect (and, therefore, must be causal), K-K relations and their derivatives can be used to place some realistic constraints on the properties themselves. The K-K relations have, of late, been used as a tempering check on the optimism that has emerged in the area of metamaterials research. This causality check includes, on one end of the spectrum, placing some realistic constraints on the application potential of metamaterials as cloaking devices \cite{stockman2007criterion} to, on the other end, sobering realizations that a considerable amount of metamaterials research stands on shaky foundations, often proposing materials which violate such basic ideas as causality and/or the second law of thermodynamics \cite{simovski2009material}. This has led to a number of researchers advocating a need for improved models for metamaterials \cite{simovski2007local,silveirinha2008spatial,silveirinha2009nonlocal,alu2011first}.

Closely connected to the idea of causality is the concept of passivity which refers to the assumption that the physical process under consideration cannot produce energy \cite{bernland2012integral}. In fact, if the physical process (cause-effect relationship) can be expressed in a convolution form in the time domain then its satisfaction of the passivity requirement automatically means that it satisfies causality as well \cite{zemanian1963n}. A physical process can, in turn, be expressed in the convolution form if it satisfies certain conditions such as linearity and time-invariance. It becomes interesting, therefore, to understand what constraints are placed upon a linear time-invariant cause-effect relationship (constitutive relationship) in electromagnetic, acoustic, and/or elastodynamic areas by the requirement of passivity. Such knowledge can be used to place constraints on and understand the limitations of various metamaterial models which are used in these areas to arrive at such relationships. Considerable research in this direction has already taken place in the field of electromagnetics where it is clear that passivity demands that the imaginary parts of the diagonal values of the Fourier transform of $\boldsymbol{\epsilon},\boldsymbol{\mu}$ be non-negative for all positive values of frequency\cite{schwinger1998classical,tip1998linear,tip2004linear,gralak2009macroscopic,liu2013causality} (fields assumed to depend upon $e^{-i\omega t}$). However, it is not clear, to the author's knowledge, what should be the equivalent constraints in elastodynamics for the most general constitutive cases. The case of 1-D longitudinal or shear wave propagation in an elastic medium is equivalent, in form, to the electromagnetic case. As such, it immediately follows that passivity should require that the corresponding 1-D material properties (modulus and density) should behave analogously to the $\epsilon,\mu$. However, in 2- and 3-D, the elastodynamic constitutive tensor cannot, in general, be diagonalized. Moreover, recent advancements \cite{milton2007modifications,nemat2011homogenization,shuvalov2011effective,willis2011effective,willis2012construction,srivastava2012overall,nemat2011overall,norris2011elastic,norris2012analytical,parnell2013antiplane} suggest that the Willis constitutive relation\cite{willis1997dynamics}, which is a coupled form of constitutive relation, is more appropriate for the description of inhomogeneous elastodynamics and, therefore, of elastodynamic metamaterials. It is not clear what the constraints of passivity are on such highly general elastodynamic constitutive forms. 

In this paper we study the constraints which passivity places on highly general forms of the elastodynamic constitutive relations. We use the passivity condition which is equivalent to the statement of passivity used in electromagnetism \cite{gralak2009macroscopic} and circuit theory \cite{zemanian1963n} and which is elaborated in subsequent sections. We also present our analysis within the context of distributions which is the proper space within which to describe the transfer functions of passive systems. Furthermore, treating the constitutive tensors in the space of distributions ensures that the analysis applies to the metamaterial cases of most interest and also to the static case (elastic case). 

\section{Background}
Physical processes in the real world are often described as an interplay between physical variables and fields which are dependent upon time. The relationships between the physical variables can be modeled as input-output relations where a time dependent variable $v(t)$ is produced from another time dependent variable $u(t)$ through some rule $\mathfrak{R}$, $v(t)=\mathfrak{R}u(t)$. Although physical variables often satisfy certain continuity and differentiability conditions, it is desirable to consider them as generalized functions for broader applicability. We identify four spaces at this point. Space $\mathcal{D}$ is the space of all complex valued functions $\phi(t)$ which are infinitely smooth and with compact support. It is a subset of space $\mathcal{S}$ which consists of all infinitely smooth and complex-valued functions $\phi(t)$, called functions of rapid descent, such that they and all their derivatives decrease to zero faster than every power of $1/|t|$ as $|t|\rightarrow\infty$. Space $\mathcal{S}^{'}$ of distributions of slow growth is the space of continuous linear functionals on $\mathcal{S}$. Space $\mathcal{D}^{'}$  is the space of continuous linear functionals on $\mathcal{D}$. It can be shown that $\mathcal{D}\subset\mathcal{S}\subset\mathcal{S}^{'}\subset\mathcal{D}^{'}$. Input-output relations can be completely arbitrary but they reduce to a particularly simple form if the properties of single-valuedness, linearity, continuity, and time-translational invariance are assumed to hold for the operator $\mathfrak{R}$ \cite{zemanian1965distribution}. These conditions are generally true for physical processes and under these conditions the operator $\mathfrak{R}$ reduces to a convolution operation:
\begin{equation}
v(t)=R*u=\int_\mathbb{R}\mathrm{d}\tau\;R(t-\tau)u(\tau)
\end{equation}
where the last equality only holds if $R,u$ are locally integrable distributions whose supports satisfy certain boundedness properties (either $R,u$ have bounded supports, or both $R,u$ are either bounded on the left or on the right). The operator $R$ is causal if it is not supported on $t<0$. The final property of passivity can be stated by defining the energy of the system. If the power absorbed by the system at time $s$ is given by Re $v^*(s)u(s)$ where $*$ denotes the complex conjugate, then define the energy absorbed by the system up to time $t$ as:
\begin{equation}\label{passivityOp}
\mathcal{E}(t)=\mathrm{Re}\int_{-\infty}^{t}\mathrm{d}s\;v^*(s)u(s)
\end{equation}
The operator $R$ is considered passive if $\mathcal{E}(t)\geq 0$ for all $t\in\mathbb{R}$. For operators in convolution passivity implies causality and it also implies that $R\in\mathcal{S}^{'}$. Therefore, Eq. (\ref{passivityOp}) is well defined for at least for all $u\in\mathcal{S}$. For distributions in $\mathcal{S}^{'}$ the Fourier transform, denoted by $\mathcal{F}$, must satisfy $\langle\mathcal{F}f,\phi\rangle=\langle f,\mathcal{F}\phi\rangle$ for $f\in\mathcal{S}^{'},\phi\in\mathcal{S}$ where $\langle f,\phi\rangle$ is the value in $\mathbb{C}$ that $f$ assigns to $\phi$ through the operation $\int_{-\infty}^\infty f(t)\phi(t)\mathrm{d}t$. Furthermore it can be shown that the Fourier transform of a distribution which is in $\mathcal{S}^{'}$ is itself in $\mathcal{S}^{'}$. For distributions in $\mathcal{S}$ the Fourier transform is defined in the usual way ($\omega\in\mathbb{R}$):
\begin{eqnarray}
\displaystyle \mathcal{F}\phi\equiv\tilde{\phi}(\omega)=\int_\mathbb{R}\mathrm{d}t\;{\phi}(t)e^{i\omega t};\quad  \phi(t)=\frac{1}{2\pi}\int_\mathbb{R}\mathrm{d}\omega\;\tilde{\phi}(\omega)e^{-i\omega t}
\end{eqnarray}
In addition to $R$ being a distribution of slow growth, causality implies that its support is in $[0,\infty)$. For such a case its Laplace transform is given by $\hat{R}=\langle R,e^{-zt}\rangle$ and it can be shown that passivity directly implies that $\mathrm{Re}\;\hat{R}\geq 0$ for all $\mathrm{Re}\;z>0$ (passivity constraint) and that $\hat{R}$ is analytic there \cite{konig1958lineare}. It is important to point out here that if $R$ is causal and is in $\mathcal{S}^{'}$ then its Laplace transform can be derived from its Fourier transform through its analytic extension in the right half of the complex plane. This is the point of departure for deriving K-K relations in electromagnetism when the constitutive tensors can be expressed diagonally. 

\textbf{For transfer functions of higher order} and more complexity which form the object of study of this paper, we need to define some additional spaces. We will use bold symbols to denote tensors whose elements are distributions. If $\mathbf{f}(t)$ is a tensor of distributions, $\langle\mathbf{f}(t),\phi(t)\rangle$ is the matrix of complex numbers obtained by replacing each element of $\mathbf{f}(t)$ by the number that this element assigns to the testing function $\phi(t)$. We will use additional subscripts with the spaces already defined above to denote the space in which all tensors of the relevant rank and distribution lie. For e.g. $\mathcal{D}^{'}_{n\times n\times n\times n}$ is the space of all fourth order tensors whose elements are distributions in $\mathcal{D}^{'}$ etc. For a tensor of even rank $\mathbf{f}$ we also define the operations $\mathbf{f}^T$ which denotes a transpose over the major symmetry and $\mathbf{f}^\dagger$ which denotes a transpose over the major symmetry followed by conjugation. Now a single-valued, linear, time-invariant, and continuous input output relation can be written in the convolution form:
\begin{equation}
\mathbf{v}(t)=\mathbf{R}*\mathbf{u}=\int_\mathbb{R}\mathrm{d}\tau\;\mathbf{R}(t-\tau)\mathbf{u}(\tau)
\end{equation}
where appropriate boundedness of $\mathbf{R},\mathbf{u}$ are assumed and where $\mathbf{v}$ is a tensorial quantity derived from $\mathbf{u}$ through the linear operator $\mathbf{R}$. Total energy absorbed up to time $t$ is given by:
\begin{equation}\label{passivityOpT}
\mathcal{E}(t)=\mathrm{Re}\int_{-\infty}^{t}\mathrm{d}s\;\mathbf{v}^\dagger(s)\mathbf{u}(s)
\end{equation}
The operator $\mathbf{R}$ is considered passive if $\mathcal{E}(t)\geq 0$ for all $t\in\mathbb{R}$ and to make sure that the integral above exists, we will restrict the elements of the input $\mathbf{u}$ to be in $\mathcal{D}$. If the operator $\mathbf{R}$ which is in the convolution form is passive then it can also be shown to be causal \cite{zemanian1963n} and, furthermore, its elements are in $\mathcal{S}^{'}$. Moreover, certain important properties of the Laplace and Fourier transform of $\mathbf{R}$ follow. Its Laplace transform is given by $\hat{\mathbf{R}}=\langle\mathbf{R},e^{-zt}\rangle$ and it is analytic (its elements are analytic) in the open right half of the complex plane. It can be derived from its Fourier transform through analytic continuation in its region of convergence. And finally, its hermitian part ($\hat{\mathbf{R}}^h=.5\left[\hat{\mathbf{R}}+\hat{\mathbf{R}}^\dagger\right]$) is positive semi-definite for all $\mathrm{Re}\;z>0$. Note that these results follow from some fairly unrestrictive constraints on the input field and transfer function which are easily satisfied in elastodynamics (and electromagnetism). Our effort here is to apply and extend these results to the elastodynamic case.

\section{Causality and passivity in elastodynamics}
We begin by considering a volume $\Omega$ within which the pointwise elastodynamic equation of motion and kinematic relations are specified:
\begin{equation}\label{equationofmotion}
\sigma_{ij,j}+f_i=\dot{p}_i; \quad \varepsilon_{ij}=\frac{1}{2}(u_{i,j}+u_{j,i}),
\end{equation}
where $\boldsymbol{\sigma},\boldsymbol{\varepsilon},\mathbf{p},\mathbf{u}$ and $\mathbf{f}$ are the space and time dependent stress tensor, strain tensor, momentum vector, displacement vector, and body force vector respectively. These relations need to be supplied with appropriate constitutive relations which relate the various field variables to each other. For the current discussion we consider stress and velocity to be independent fields (input fields) which lead to the emergence of strain and momentum fields (output/dependent fields) respectively. The relationships are expressed in terms of general constitutive operators whose properties need to be determined based upon the various subsequent assumptions about the system:
\begin{equation}\label{constitutiveOp}
\boldsymbol{\epsilon}(\mathbf{x},t)=\mathfrak{D}(\boldsymbol{\sigma}(\mathbf{x},t)), \quad \mathbf{p}(\mathbf{x},t)=\mathfrak{P}(\dot{\mathbf{u}}(\mathbf{x},t))
\end{equation}
Now we assume that the operators satisfy the conditions of single-valuedness, linearity, time-invariance, and continuity and, furthermore, that the stress and velocity fields along with the operators satisfy appropriate boundedness conditions referred to in the last section. We also assume that the operator is real valued, i.e. it assigns a real output field to a real input field (but can assign a complex output field to a complex input field). Under these assumptions the constitutive relations of Eq. (\ref{constitutiveOp}) can be specialized to the following form:
\begin{eqnarray}\label{constitutive}
\displaystyle \nonumber \boldsymbol{\epsilon}(\mathbf{x},t)=\mathbf{D}*\boldsymbol{\sigma}=\int_{-\infty}^{\infty}\mathrm{d}s\; \mathbf{D}(\mathbf{x},t-s):\boldsymbol{\sigma}(\mathbf{x},s)=\int_{-\infty}^{\infty}\mathrm{d}s\; \mathbf{D}(\mathbf{x},s):\boldsymbol{\sigma}(\mathbf{x},t-s)\\
\displaystyle \mathbf{p}(\mathbf{x},t)=\boldsymbol{\rho}*\dot{\mathbf{u}}=\int_{-\infty}^{\infty}\mathrm{d}s\;\boldsymbol{\rho}(\mathbf{x},t-s)\cdot\dot{\mathbf{u}}(\mathbf{x},s)=\int_{-\infty}^{\infty}\mathrm{d}s\;\boldsymbol{\rho}(\mathbf{x},s)\cdot\dot{\mathbf{u}}(\mathbf{x},t-s)
\end{eqnarray}
where the components of the constitutive tensors $\mathbf{D},\boldsymbol{\rho}$ are real valued distributions. $\mathbf{D}$ is a fourth order tensor field in $\mathcal{S}^{'}_{n\times n\times n\times n}$ and possesses the usual minor symmetries that are associated with stiffness and compliance tensors. $\boldsymbol{\rho}$ is a second order tensor field in $\mathcal{S}^{'}_{n\times n}$. To make sure that energy, as defined later on, exists, we will take $\boldsymbol{\sigma}$ to be in $\mathcal{D}_{n\times n}$ and $\dot{\mathbf{u}}$ to be in $\mathcal{D}_{n\times 1}$. No major symmetries are assumed at this point. We also note a further result which will be used later. For convolutions of distributions as appearing above we note that $\dot{\boldsymbol{\epsilon}}=\dot{\mathbf{D}}*\boldsymbol{\sigma}$ and $\dot{\mathbf{p}}=\dot{\boldsymbol{\rho}}*\dot{\mathbf{u}}$.

\textbf{Causality}: Causality refers to the requirement that an effect cannot precede its cause. With reference to the constitutive relations in Eq. (\ref{constitutive}), it implies that the value of the strain and momentum fields at time $t_0$ can only depend upon the values, respectively, of the stress and velocity fields at times prior to and including $t_0$. A necessary and sufficient condition for a system to be causal is that its unit response function (constitutive operator in the present case) vanishes for $t<0$. Specifically, causality implies the following for the constitutive tensors:
\begin{equation}\label{causality}
\mathbf{D}(\mathbf{x},t)=0,\;\; \boldsymbol{\rho}(\mathbf{x},t)=0 \;\;\forall t<0
\end{equation}

\textbf{Passivity}: Passivity refers to the requirement that the system cannot generate energy. For the elastodynamic case the total energy at any time $t$ contained in $\Omega$ comprises of the elastic energy contribution and the kinetic energy contribution:
\begin{equation}\label{energy}
\mathcal{E}(t)=\mathrm{Re}\;\frac{1}{2}\int_\Omega\mathrm{d}\mathbf{x}\;\left[\boldsymbol{\sigma}(\mathbf{x},t):\boldsymbol{\epsilon}^*(\mathbf{x},t)+\mathbf{p}(\mathbf{x},t)\cdot\dot{\mathbf{u}}^*(\mathbf{x},t)\right]
\end{equation}
The total energy can, therefore, be given by the following integral:
\begin{equation}\label{energyAbs}
\mathcal{E}(t)=\int_{-\infty}^t\mathrm{d}s\;\frac{\partial\mathcal{E}(s)}{\partial s}=\mathrm{Re}\;\frac{1}{2}\int_{-\infty}^{t}\mathrm{d}s\;\int_\Omega\mathrm{d}\mathbf{x}\;\frac{\partial}{\partial s}\left[\boldsymbol{\sigma}(\mathbf{x},s):\boldsymbol{\epsilon}^*(\mathbf{x},s)+\mathbf{p}(\mathbf{x},s)\cdot\dot{\mathbf{u}}^*(\mathbf{x},s)\right]
\end{equation}
Passivity requires that the total energy absorbed by the system be non-negative at all times:
\begin{equation}\label{passivity}
\mathcal{E}(t)\geq 0\;\;\forall\;t
\end{equation}
To calculate the total energy absorbed by the system in Eq. (\ref{energyAbs}), we note that the time derivative of $\mathcal{E}(t)$ should equal the power input from the tractions, $\mathbf{t}(\mathbf{x},t)$, which are acting on $\partial\Omega$ and the body forces, $\mathbf{f}$, which are acting in $\Omega$ \cite{achenbach1984wave}. This power input is given by:
\begin{equation}\label{power}
P(t)=\mathrm{Re}\;\left[\int_{\partial\Omega}\mathrm{d}\mathbf{x}\; t_i(\mathbf{x},t)\dot{u}_i^*(\mathbf{x},t) +\int_{\Omega} \mathrm{d}\mathbf{x}\;f_i(\mathbf{x},t)\dot{u}_i^*(\mathbf{x},t)\right]\equiv\mathrm{Re}\;\int_{\Omega} \mathrm{d}\mathbf{x}\;\left[(\sigma_{ij}\dot{u}_i^*)_{,j}+f_i\dot{u}_i^*\right] 
\end{equation}
where the dependence on space and time is suppresed. The above is achieved through the application of the Gauss theorem and the relation $t_i=\sigma_{ij}n_j$. By decomposing $\dot{u}_{i,j}=\dot{\epsilon}_{ij}+\dot{\omega}_{ij}$ where $\dot{\epsilon}$ and $\dot{\omega}$ are the symmetric and antisymmetric parts of $\dot{u}_{i,j}$, respectively, and noting that the inner product of a symmetric tensor and an antisymmetric tensor goes to zero, we have:
\begin{equation}\label{power1}
P(t)=\mathrm{Re}\;\int_{\Omega}\mathrm{d}\mathbf{x}\;\left[\sigma_{ij,j}\dot{u}_i^*+\sigma_{ij}\dot{\epsilon}_{ij}^*+f_i\dot{u}_i^*\right] 
\end{equation}
Using the equation of motion (Eq. \ref{equationofmotion}) and rearranging, we have:
\begin{equation}\label{power2}
P(t)=\mathrm{Re}\;\int_{\Omega} \mathrm{d}\mathbf{x}\;\left[\sigma_{ij}\dot{\epsilon}_{ij}^*+\dot{p}_i\dot{u}_i^*\right] 
\end{equation}
The above calculated power should equal the rate of change of the total energy stored in $\Omega$:
\begin{equation}\label{powerenergy}
\frac{d\mathcal{E}(t)}{dt}=P(t)
\end{equation}
Eqs. (\ref{energyAbs},\ref{passivity},\ref{powerenergy}) together give:
\begin{equation}\label{passivity2}
\mathcal{E}(t)=\int_{-\infty}^{t}\mathrm{d}s\;P(s)=\mathrm{Re}\;\int_{-\infty}^{t}\mathrm{d}s\;\int_\Omega \mathrm{d}\mathbf{x}\left[\sigma_{ij}(\mathbf{x},s)\dot{\epsilon}_{ij}^*(\mathbf{x},s)+\dot{p}_i(\mathbf{x},s)\dot{u}_i^*(\mathbf{x},s)\right]\geq 0
\end{equation}
Since $\Omega$ is arbitrary, the above would be satisfied only if the inequality holds at each point in space. Going forward we, therefore, understand the passivity statement to be the following:
\begin{equation}\label{passivity1}
\mathrm{Re}\;\int_{-\infty}^{t}\mathrm{d}s\;\left[\sigma_{ij}(s)\dot{\epsilon}_{ij}^*(s)+\dot{p}_i(s)\dot{u}_i^*(s)\right]\geq 0
\end{equation}
where the variables are being evaluated at an arbitrary but same location $\mathbf{x}\in\Omega$. Since the constitutive tensors in Eq. (\ref{constitutive}) are real we note that $\dot{\boldsymbol{\epsilon}}^*=\dot{\mathbf{D}}*\boldsymbol{\sigma}^*$ and $\dot{\mathbf{p}}=\dot{\boldsymbol{\rho}}*\dot{\mathbf{u}}$. Using these relations we have:
\begin{equation}
\mathrm{Re}\;\int_{-\infty}^{t}\mathrm{d}s\;\left[\sigma_{ij}(s)\int_{-\infty}^{\infty}\mathrm{d}v\; \dot{D}_{ijkl}(v)\sigma_{kl}^*(s-v)+\dot{u}_i^*(s)\int_{-\infty}^{\infty}\mathrm{d}v\;\dot{\rho}_{ij}(v)\dot{u}_j(s-v)\right]\geq 0
\end{equation}
We note that the time derivatives of the constitutive tensors should themselves be real and causal and, therefore, must satisfy $\dot{\mathbf{D}}(\mathbf{x},t)=\dot{\boldsymbol{\rho}}(\mathbf{x},t)=0$ for all $t<0$. Now we employ the distributional Laplace transform. This is done by first choosing $\boldsymbol{\sigma}(s)=\boldsymbol{\sigma}\phi^*(s)$ and $\dot{\mathbf{u}}(s)=\dot{\mathbf{u}}\gamma(s)$ where $\boldsymbol{\sigma},\dot{\mathbf{u}}$ are constant tensors and $\phi(s),\gamma(s)$ are in $\mathcal{D}$:
\begin{equation}\label{passivity3}
\mathrm{Re}\;\int_{-\infty}^{t}\mathrm{d}s\;\left[\sigma_{ij}\phi^*(s)\int_{-\infty}^{\infty}\mathrm{d}v\; \dot{D}_{ijkl}(v)\sigma_{kl}^*\phi(s-v)+\dot{u}_i^*\gamma^*(s)\int_{-\infty}^{\infty}\mathrm{d}v\;\dot{\rho}_{ij}(v)\dot{u}_j\gamma(s-v)\right]\geq 0
\end{equation}
and then letting $\phi(s),\gamma(s)$ be equal to $e^{zs}$ for $-\infty<s<a$ where $t<a<\infty$ and $z\in\mathbb{C}$. With the requirement of causality on $\dot{\mathbf{D}},\dot{\boldsymbol{\rho}}$ the above inequality reduces to the following:
\begin{equation}
\mathrm{Re}\left[\sigma_{ij}\sigma_{kl}^*\int_{0}^{\infty}\mathrm{d}v\; \dot{D}_{ijkl}(v)e^{-zv} +\dot{u}_{i}^*\dot{u}_{j}\int_{0}^{\infty}\mathrm{d}v\; \dot{\rho}_{ij}(v)e^{-zv}\right]\;\int_{-\infty}^{t}e^{2z_rs}\mathrm{d}s\geq 0
\end{equation}
or
\begin{equation}
\mathrm{Re}\left[\sigma_{ij}\sigma_{kl}^*\int_{0}^{\infty}\mathrm{d}v\; \dot{D}_{ijkl}(v)e^{-zv}+\dot{u}_{i}^*\dot{u}_{j}\int_{0}^{\infty}\mathrm{d}v\; \dot{\rho}_{ij}(v)e^{-zv}\right]\geq 0
\end{equation}
Since the constitutive tensors are distributions of slow growth, the integral quantities are identified as the distributional Laplace transforms of the respective constitutive tensors so that the above relations can be expressed in the following condensed form:
\begin{equation}\label{energyAbs1}
\mathrm{Re}\left[\boldsymbol{\sigma}:\hat{\dot{\mathbf{D}}}:\boldsymbol{\sigma}^*+\dot{\mathbf{u}}\cdot\hat{\dot{\boldsymbol{\rho}}}\cdot\dot{\mathbf{u}}^*\right]\geq 0
\end{equation}
We now decompose the tensors $\hat{\dot{\mathbf{D}}}$ and $\hat{\dot{\boldsymbol{\rho}}}$ into their hermitian and non-hermitian parts:
\begin{eqnarray}\label{hermitian}
\displaystyle \nonumber \hat{\dot{\mathbf{D}}}=\hat{\dot{\mathbf{D}}}^h+\hat{\dot{\mathbf{D}}}^{nh};\quad \hat{\dot{\mathbf{D}}}^h=\frac{1}{2}\left[\hat{\dot{\mathbf{D}}}+\hat{\dot{\mathbf{D}}}^{\dagger}\right];\quad \hat{\dot{\mathbf{D}}}^{nh}=\frac{1}{2}\left[\hat{\dot{\mathbf{D}}}-\hat{\dot{\mathbf{D}}}^{\dagger}\right]\\
\hat{\dot{\boldsymbol{\rho}}}=\hat{\dot{\boldsymbol{\rho}}}^h+\hat{\dot{\boldsymbol{\rho}}}^{nh};\quad \hat{\dot{\boldsymbol{\rho}}}^h=\frac{1}{2}\left[\hat{\dot{\boldsymbol{\rho}}}+\hat{\dot{\boldsymbol{\rho}}}^{\dagger}\right];\quad \hat{\dot{\boldsymbol{\rho}}}^{nh}=\frac{1}{2}\left[\hat{\dot{\boldsymbol{\rho}}}-\hat{\dot{\boldsymbol{\rho}}}^{\dagger}\right]
\end{eqnarray}
and note that only hermitian parts of the tensors contribute to the real part of Eq. (\ref{energyAbs1}). Furthermore, since the tensors ${\boldsymbol{\sigma}},{\dot{\mathbf{u}}}$ in Eq. (\ref{energyAbs1}) are arbitrary the inequality $\mathcal{E}(t)\geq 0$ implies the positive semi-definiteness of the hermitian parts of the constitutive tensors in the Laplace domain:
\begin{eqnarray}\label{positivedef}
\displaystyle \nonumber \boldsymbol{\phi}:\hat{\dot{\mathbf{D}}}^h:\boldsymbol{\phi}^*\geq 0\\
\displaystyle \mathbf{q}\cdot\hat{\dot{\boldsymbol{\rho}}}^h\cdot\mathbf{q}^*\geq 0
\end{eqnarray}
where $\boldsymbol{\phi}$ is an arbitrary complex-valued second order symmetric tensor, $\mathbf{q}$ is an arbitrary complex-valued vector, and the relation holds for all $\mathbf{x}$ and $z$ (in the region of convergence). Similar results can be derived for the Fourier transform of the time derivatives of the constitutive tensors. Under the restriction that the support of a distribution $\mathbf{f}$ be bounded, its Fourier transform is given by $\langle\mathbf{f}(t),e^{i\omega t}\rangle$. Now we let $\phi(t),\gamma(t)$ be equal to $e^{-i\omega t}$ in Eq. (\ref{passivity3}) and follow the subsequent process to determine that the hermitian parts of the Fourier transforms of the time-derivative constitutive tensors must be positive semi-definite. However, the boundedness restrictions on the constitutive tensors need not be so severe for us to come to this conclusion. We merely assume that the constitutive tensors are distributions of slow growth to come to the same conclusion. To do so we consider the following for a test function $\phi(t)\in\mathcal{S}$ and a distribution $f(t)\in\mathcal{S}^{'}$:
\begin{eqnarray}
\displaystyle \nonumber
\int_{-\infty}^{t}\mathrm{d}s\phi^*(s)\int_{-\infty}^{\infty}\mathrm{d}vf(v)\phi(s-v)=\frac{1}{2\pi}\int_{-\infty}^{t}\mathrm{d}s\phi^*(s)\int_{-\infty}^{\infty}\mathrm{d}\omega \tilde{f}(\omega)\int_{-\infty}^{\infty}\phi(s-v)e^{-i\omega v}\mathrm{d}v\\
\displaystyle =\frac{1}{2\pi}\int_{-\infty}^{\infty}\mathrm{d}\omega\int_{-\infty}^{t}\mathrm{d}s\phi^*(s)e^{-i\omega s} \tilde{f}(\omega)\int_{-\infty}^{\infty}\mathrm{d}u\phi(u)e^{i\omega u}=\frac{1}{2\pi}\int_{-\infty}^{\infty}\mathrm{d}\omega\bar{\phi}(\omega) \tilde{f}(\omega)\bar{\phi}^*(\omega)
\end{eqnarray}
where the distributional Fourier transform is used and the last step follows by choosing the test function $\phi(\tau)$ such that it vanishes for $\tau> t$. It is clear from the above that under the much less restrictive conditions that the constitutive tensors be distributions of slow growth, Eq. (\ref{passivity3}) can be written, after some manipulations, in the following way:
\begin{equation}\label{energyAbs2}
\mathrm{Re}\int_{-\infty}^{\infty}\mathrm{d}\omega\left[\boldsymbol{\sigma}(\omega):\tilde{\dot{\mathbf{D}}}(\omega):\boldsymbol{\sigma}^*(\omega)+\dot{\mathbf{u}}(\omega)\cdot\tilde{\dot{\boldsymbol{\rho}}}(\omega)\cdot\dot{\mathbf{u}}^*(\omega)\right]\geq 0
\end{equation}
Since ${\boldsymbol{\sigma}},{\dot{\mathbf{u}}}$ are arbitrary, the above will be satisfied only if the following is true about the hermitian parts of the Fourier transforms:
\begin{eqnarray}\label{positivedefF}
\displaystyle \nonumber \boldsymbol{\phi}:\tilde{\dot{\mathbf{D}}}^h:\boldsymbol{\phi}^*\geq 0\\
\displaystyle \mathbf{q}\cdot\tilde{\dot{\boldsymbol{\rho}}}^h\cdot\mathbf{q}^*\geq 0
\end{eqnarray}
Eqs. (\ref{positivedef},\ref{positivedefF}) constitute important conclusions in this paper and they are in line with earlier conclusions in circuit theory \cite{zemanian1972realizability}. These relations can be used to place similar constraints on the transforms of constitutive tensors $\mathbf{D},\boldsymbol{\rho}$. For this we need only consider the relations between the transforms of derivates as they apply to distributions in $\mathcal{S}^{'}$. The following relations are noted:
\begin{equation}
\tilde{\dot{f}}=-i\omega\tilde{{f}};\quad \hat{\dot{f}}=z\hat{{f}};\quad f\in\mathcal{S}^{'}
\end{equation}
Since the elements of our constitutive tensors are assumed to be in $\mathcal{S}^{'}$ the above relations apply to them. Specifically we have, for instance for the Fourier transforms:
\begin{eqnarray}\label{DTimeDerFourier}
\displaystyle \tilde{\mathbf{D}}(\omega)=-\frac{\tilde{\dot{\mathbf{D}}}(\mathbf{x},\omega)}{i\omega}=\frac{i\tilde{\dot{\mathbf{D}}}(\mathbf{x},\omega)}{\omega};\quad \tilde{\boldsymbol{\rho}}(\mathbf{x},\omega)=\frac{i\tilde{\dot{\boldsymbol{\rho}}}(\mathbf{x},\omega)}{\omega}
\end{eqnarray}
Eq. (\ref{DTimeDerFourier}) explicitly relates the Fourier transforms of the constitutive tensors to the Fourier transforms of their time derivatives. The Laplace transforms can be similarly related. To extend the constraints which passivity applies to the Fourier transforms of the constitutive tensors (analogous to Eq. \ref{positivedefF}) we use the hermitian transpose operation, $\dagger$. It is clear that $\tilde{{\mathbf{D}}}^{h\dagger}=\mathbf{D}^{h}$ and $\tilde{{\mathbf{D}}}^{nh\dagger}=-\mathbf{D}^{nh}$ with similar relations holding for second order tensors. By expanding $\tilde{\mathbf{D}},\tilde{\boldsymbol{\rho}},\tilde{\dot{\mathbf{D}}},\tilde{\dot{\boldsymbol{\rho}}}$ into their hermitian and non-hermitial parts in Eq. (\ref{DTimeDerFourier}) and by applying the hermitian transpose operator it becomes clear that the hermitian parts of the time derivative quantities are related to the non-hermitian parts of the original tensors in the following sense ($\omega$ dependent implicit):
\begin{eqnarray}
\displaystyle \tilde{\dot{\mathbf{D}}}^h=\frac{\omega}{i}\tilde{\mathbf{D}}^{nh};\quad \tilde{\dot{\boldsymbol{\rho}}}^h=\frac{\omega}{i}\tilde{\boldsymbol{\rho}}^{nh}
\end{eqnarray}
Therefore, now Eq. (\ref{positivedefF}) places the following constraints on the Fourier transforms of $\mathbf{D},\boldsymbol{\rho}$:
\begin{eqnarray}\label{positivedefO}
\displaystyle \nonumber \boldsymbol{\phi}:\frac{\omega}{i}\tilde{\mathbf{D}}^{nh}(\mathbf{x},\omega):\boldsymbol{\phi}^*\geq 0\\
\displaystyle \mathbf{q}\cdot\frac{\omega}{i}\tilde{\boldsymbol{\rho}}^{nh}(\mathbf{x},\omega)\cdot\mathbf{q}^*\geq 0
\end{eqnarray}
It is clear that operations such as $\boldsymbol{\phi}:\tilde{\mathbf{D}}^{nh}(\mathbf{x},\omega):\boldsymbol{\phi}^*$ and $\mathbf{q}\cdot\tilde{\boldsymbol{\rho}}^{nh}(\mathbf{x},\omega)\cdot\mathbf{q}^*$ result in purely imaginary numbers. However the factor $i$ in the denominator ensures that the quantities in Eq. (\ref{positivedefO}) are purely real. Passivity and causality of the system demand that these numbers also be non-negative for positive values of $\omega$ (and non-positive for negative values of $\omega$). We also note the corollary result that had we decided to represent Fourier transform through the exponential $e^{-i\omega t}$ instead of $e^{i\omega t}$ we would have arrived at the complementary result where the non-hermitian quantities above would have been required to be negative semi-definite instead of positive semi-definite. In the following sections we will consider a specialization and a generalization of the above results. The specialization refers to cases where the constitutive tensors possess major symmetries and the generalization refers to the above results in the context of more general forms of constitutive relations such as the Willis kind of coupled relations.

\section{With Major Symmetries}

Up to now we have assumed no special forms for the compliance and density tensors beyond the minor symmetries which ensure rotational stability of the system. We now consider the specialization of the above results to a case where the constitutive tensors possess major symmetries as well. For the density tensor we mean that its components satisfy $\rho_{ij}=\rho_{ji}$. Similarly we require that the fourth order compliance tensor $D_{ijkl}$ satisfy $D_{ijkl}=D_{klij}$. Since the components of the Fourier and Laplace transforms of the constitutive tensors are only related to the corresponding time domain components, it is clear that these major symmetries will extend to them as well. With these additional requirements Eqs. (\ref{hermitian}) imply that $\tilde{\dot{\mathbf{D}}}^{h*}=\tilde{\dot{\mathbf{D}}}^{h}$ and $\tilde{\dot{\boldsymbol{\rho}}}^{h*}=\tilde{\dot{\boldsymbol{\rho}}}^{h}$ essentially meaning that $\tilde{\dot{\mathbf{D}}}^{h}$ and $\tilde{\dot{\boldsymbol{\rho}}}^{h}$ are composed only of the real parts of $\tilde{\dot{\mathbf{D}}}$ and $\tilde{\dot{\boldsymbol{\rho}}}$ respectively. Similarly, $\tilde{\dot{\mathbf{D}}}^{nh}$ and $\tilde{\dot{\boldsymbol{\rho}}}^{nh}$ are composed only of the imaginary parts of $\tilde{\dot{\mathbf{D}}}$ and $\tilde{\dot{\boldsymbol{\rho}}}$ respectively. Furthermore, these characteristics should hold for all tensors of current interest (i.e. Fourier and Laplace transforms of $\mathbf{D},\boldsymbol{\rho}$ as well). Consideration of this specialization is of interest because for this case, the definiteness relations apply simply to the real and imaginary parts of the relevant tensors. Specifically, we have the following relations for this case:
\begin{eqnarray}\label{positivedefS}
\displaystyle \nonumber \boldsymbol{\phi}:\mathrm{Re}\;\tilde{\dot{\mathbf{D}}}:\boldsymbol{\phi}^*\geq 0;\quad \mathbf{q}\cdot\mathrm{Re}\;\tilde{\dot{\boldsymbol{\rho}}}\cdot\mathbf{q}^*\geq 0\quad\forall \boldsymbol{\phi}=\boldsymbol{\phi}^T,\mathbf{q}\\
\displaystyle \boldsymbol{\phi}:\frac{\omega}{i}\mathrm{Im}\;\tilde{{\mathbf{D}}}:\boldsymbol{\phi}^*\geq 0;\quad \mathbf{q}\cdot\frac{\omega}{i}\mathrm{Im}\;\tilde{{\boldsymbol{\rho}}}\cdot\mathbf{q}^*\geq 0\quad\forall \boldsymbol{\phi}=\boldsymbol{\phi}^T,\mathbf{q}
\end{eqnarray}
Restricting our attention for this section to the real parts (denoted by the subscript r) of the time derivative, fourier transformed tensors and to the imaginary parts (denoted by the subscript i) of the fourier transformed tensors, and using the shorthand $\geq$ to imply positive semi-definiteness, the above relations are condensed to ($\omega$ dependence implied):
\begin{eqnarray}
\displaystyle \nonumber \tilde{\dot{\mathbf{D}}}_r,\tilde{\dot{\boldsymbol{\rho}}}_r,\omega\tilde{{\mathbf{D}}}_i,\omega\tilde{{\boldsymbol{\rho}}}_i\geq 0 \quad \forall\; \mathbf{x},\omega
\end{eqnarray}
Similar relations are derived by Milton and Willis \cite{milton2010minimum} in the context of minimum variational principles for time-harmonic waves in a dissipative medium. Since the imaginary parts of the Fourier transformed constitutive relations, for the simpler major-symmetric case as this one, corresponds to the dissipation in the system we note an interesting result as a corollary. \emph{For constitutive relations which do not necessarily possess major-symmetry, it is the non-hermitian parts of the Fourier transformed tensors which corresponds to dissipation in the system. In other words, a conservative system can be expected to be hermitian in the Fourier transform of its constitutive relations}. One system which immediately corresponds to the major-symmetric specialization being considered here is the case of one dimensional elastodynamics:
\begin{eqnarray}\label{constitutiveM1D}
\displaystyle \nonumber {\epsilon}({x},t)=\int_{-\infty}^{t}\mathrm{d}s\; {D}({x},t-s){\sigma}({x},s)\quad {p}({x},t)=\int_{-\infty}^{t}\mathrm{d}s\;{\rho}({x},t-s)\dot{{u}}({x},s)
\end{eqnarray}
Causality and passivity results from the earlier sections can be immediately extended here to conclude that $\hat{D},\hat{\rho}$ are analytic for Re $z>0$ and that $\hat{\dot{{D}}}_r,\hat{\dot{{\rho}}}_r,\tilde{\dot{{D}}}_r,\tilde{\dot{{\rho}}}_r,\omega\tilde{{{D}}}_i,\omega\tilde{{{\rho}}}_i\geq 0$ where the symbol $\geq$ actually means greater than or equal to in the present case and not just positive semi-definiteness. The static case, for which the constitutive tensors can be represented through the delta distribution, is seen to trivially satisfy these conditions since $\tilde{\delta}=\hat{\delta}=1$. This ensures that $\omega\tilde{{{D}}}_i,\omega\tilde{{{\rho}}}_i= 0$ etc. The symbol $\geq$ can be understood to mean greater than equal to, and not just positive semi-definiteness, whenever a diagonal constitutive relation is being considered. In those cases passivity and causality would dictate that the $\geq$ relations apply to the diagonal elements individually. Diagonal relations for the density tensor include those cases where $\rho_{ij}\propto \delta_{ij}$, and for the compliance tensor include those cases where $D_{ijkl}\propto\delta_{ik}\delta_{jl}$.

\section{Generalization to other constitutive relationships}

To derive the passivity relationships we required that energy could be expressed in a particular form (which it does automatically for the constitutive relations considered up to now). We will use this observation to generalize the results from the previous sections to more general constitutive relations such as the Willis relations. In the subsequent treatment we will understand the space dependence to be implicit in the sense of Eq. (\ref{power1}). Let $\mathbf{w}(t),\mathbf{v}(t)$ denote column vectors consisting of $n$ time dependent tensors. Elements of $\mathbf{w}(t)$ are assumed to be in $\mathcal{D}$ to ensure that the energy expression exists (elements of $\mathbf{v}(t)$ are also distributions but they need not be so restricted). Let $\mathbf{v}(t)$ be derivable from $\mathbf{w}(\mathbf{x},t)$ through a linear, real, time invariant, and causal relationship $\mathbf{v}=\mathbf{L}*\mathbf{w}$ where $\mathbf{L}$ is a $n\times n$ matrix of real valued tensors:
\begin{equation}
\mathbf{v}(t)=\int_{-\infty}^{\infty}\mathrm{d}s\; \mathbf{L}(t-s)\mathbf{w}(s)=\int_{-\infty}^{\infty}\mathrm{d}s\; \mathbf{L}(s)\mathbf{w}(t-s)
\end{equation}
Each element of every tensor in $\mathbf{L}$ is assumed to be a distribution of slow growth. Let us also assume that the energy absorbed by the system up to a time $t$ can be represented by:
\begin{equation}
\mathcal{E}(t)=\mathrm{Re}\int_{-\infty}^{t}\mathrm{d}s\;\mathbf{w}^\dagger(s)\dot{\mathbf{v}}(s)=\mathrm{Re}\int_{-\infty}^{t}\mathrm{d}s\; \mathbf{w}^\dagger(s)\int_{-\infty}^{\infty}\mathrm{d}v\; \dot{\mathbf{L}}(v)\mathbf{u}(s-v)
\end{equation}
then some conclusions apply to the Fourier and Laplace transforms of $\mathbf{L}$ and $\dot{\mathbf{L}}$. For instance, decomposing $\tilde{\dot{\mathbf{L}}}$ into its hermitian and non-hermitian parts $\tilde{\dot{\mathbf{L}}}=\tilde{\dot{\mathbf{L}}}^h+\tilde{\dot{\mathbf{L}}}^{nh}$ and noting that $(\mathbf{y}^\dagger \tilde{\dot{\mathbf{L}}}^h\mathbf{y})^\dagger=\mathbf{y}^\dagger \tilde{\dot{\mathbf{L}}}^h\mathbf{y}$ and, therefore, real and $(\mathbf{y}^\dagger \tilde{\dot{\mathbf{L}}}^{nh}\mathbf{y})^\dagger=-\mathbf{y}^\dagger \tilde{\dot{\mathbf{L}}}^{nh}\mathbf{y}$ and, therefore, imaginary. This means that $\mathcal{E}(t)$ emerges from $\tilde{\dot{\mathbf{L}}}^h$ (or $\hat{\dot{\mathbf{L}}}^h$). Passivity requires that absorbed energy must be non-negative at all times or that $\mathcal{E}(t)\geq 0$. This implies the following:
\begin{equation}\label{positivedefSG1}
\mathbf{y}^\dagger\tilde{\dot{\mathbf{L}}}^h\mathbf{y}\geq 0;\quad\mathbf{y}^\dagger\hat{\dot{\mathbf{L}}}^h\mathbf{y}\geq 0;\quad \mathbf{y}^\dagger\frac{\omega}{i}\tilde{{\mathbf{L}}}^{nh}\mathbf{y}\geq 0
\end{equation}
The Willis relations are a special case of the above. Specifically, they relate stress, strain, velocity, and momentum in a coupled constitutive relationship ($\mathbf{x},t$ dependence implied):
\begin{equation}
 \begin{pmatrix}
  \boldsymbol{\epsilon}\\
  \mathbf{p}
 \end{pmatrix} =
 \begin{bmatrix}
  \mathbf{D} & \mathbf{S}_1\\
  \mathbf{S}_2 & \boldsymbol{\rho}
 \end{bmatrix}*
 \begin{pmatrix}
  \boldsymbol{\sigma}\\
  \dot{\mathbf{u}}
 \end{pmatrix}
\end{equation}
Power from Eq. (\ref{power2}) can be written as:
\begin{equation}
P(t)=\mathrm{Re}\;\left[\sigma_{ij}\dot{\epsilon}_{ij}^*+\dot{p}_i\dot{u}_i^*\right] \equiv \mathrm{Re}\;\left[\sigma_{ij}^*\dot{\epsilon}_{ij}+\dot{u}_i^*\dot{p}_i\right] =\mathrm{Re}\; \mathbf{w}^\dagger\dot{\mathbf{v}}
\end{equation}
with
\begin{equation}
\mathbf{v}=
 \begin{pmatrix}
  \boldsymbol{\epsilon}\\
  \mathbf{p}
 \end{pmatrix};\quad
\mathbf{w}=
 \begin{pmatrix}
  \boldsymbol{\sigma}\\
  \dot{\mathbf{u}}
 \end{pmatrix}
\end{equation}
Since energy can now be written as 
\begin{equation}
\mathcal{E}(t)=\int_{-\infty}^{t}\mathrm{d}s\; P(s)=\mathrm{Re}\int_{-\infty}^{t}\mathrm{d}s\;\mathbf{w}^\dagger(s)\dot{\mathbf{v}}(s)
\end{equation}
we immediately have from the earlier results in this section that passivity implies that the following tensors will be positive semi-definite in the sense of Eq. (\ref{positivedefSG1}):
\begin{equation}
\tilde{\dot{\mathbf{L}}}^h=
 \begin{bmatrix}
  \tilde{\dot{\mathbf{D}}} & \tilde{\dot{\mathbf{S}}}_1\\
  \tilde{\dot{\mathbf{S}}}_2 & \tilde{\dot{\boldsymbol{\rho}}}
 \end{bmatrix}+
 \begin{bmatrix}
  \tilde{\dot{\mathbf{D}}} & \tilde{\dot{\mathbf{S}}}_1\\
  \tilde{\dot{\mathbf{S}}}_2 & \tilde{\dot{\boldsymbol{\rho}}}
 \end{bmatrix}^\dagger;\quad
\hat{\dot{\mathbf{L}}}^h=
 \begin{bmatrix}
  \hat{\dot{\mathbf{D}}} & \hat{\dot{\mathbf{S}}}_1\\
  \hat{\dot{\mathbf{S}}}_2 & \hat{\dot{\boldsymbol{\rho}}}
 \end{bmatrix}+
 \begin{bmatrix}
  \hat{\dot{\mathbf{D}}} & \hat{\dot{\mathbf{S}}}_1\\
  \hat{\dot{\mathbf{S}}}_2 & \hat{\dot{\boldsymbol{\rho}}}
 \end{bmatrix}^\dagger;\quad
\tilde{{\mathbf{L}}}^{nh}=
 \begin{bmatrix}
  \tilde{{\mathbf{D}}} & \tilde{{\mathbf{S}}}_1\\
  \tilde{{\mathbf{S}}}_2 & \tilde{{\boldsymbol{\rho}}}
 \end{bmatrix}-
 \begin{bmatrix}
  \tilde{{\mathbf{D}}} & \tilde{{\mathbf{S}}}_1\\
  \tilde{{\mathbf{S}}}_2 & \tilde{{\boldsymbol{\rho}}}
 \end{bmatrix}^\dagger
\end{equation}
It must be noted that the above results make no assumptions about the process by which Willis properties have been defined or derived or whether the relations are hermitian or not. They are merely the constraints which must be satisfied if an elastodynamic system, represented in the Willis form, is required to be causal. In fact, the question of whether the Willis constitutive relation displays self-adjointness (or are hermitian) or not has been addressed in literature. In Ref. \cite{willis2011effective} it has been shown that the property of self-adjointness is preserved at the level of the effective response. In other words, the effective relations are (not) self-adjoint if the constituting materials themselves are (not) self-adjoint. Ref. \cite{srivastava2012overall} talks about the related question of hermiticity. The Willis relations exhibit several degrees of non-uniqueness \cite{willis2012construction} and, to the author's knowledge, it is not clear if some, many, or all sets of Willis properties (that can be assigned under any given case) satisfy the kinds of causality and passivity requirements discussed in this paper. 

\section{Conclusions}
In this paper we clarify the constraints that causality and passivity place on the elastodynamic constitutive tensors. Analogous questions have been addressed in other fields but the elastodynamic case is generally more complicated due to the higher order and non-diagonal nature of its constitutive relations. Here we deal with the problem in considerable generality wherein the elements of the constitutive tensors are assumed to be generalized functions in time. The treatment and conclusions presented here, therefore, apply to metamaterial applications which often involve singular and coupled constitutive forms and also to the static limit where the constitutive tensors are in the form of delta distributions. Specifically we show that the satisfaction of passivity (and causality) directly requires that the hermitian parts, as defined later, of the transforms (Fourier and Laplace) of the time derivatives of the elastodynamic constitutive tensors be positive semi-definite. Additionally, the conditions subsequently require that the non-hermitian parts of the Fourier transforms of the constitutive tensors be positive semi-definite for positive values of frequency and negative semi-definite for negative values of frequency. We show that when major symmetries are assumed these definiteness relations apply simply to the real and imaginary parts of the relevant tensors. For diagonal and one-dimensional problems, these positive semi-definiteness relationships reduce to simple inequality relations over the real and imaginary parts. Finally we extend the results to highly general forms of constitutive relations which include the Willis inhomogeneous relations as a special case.

\section{Data Accessibility}
No data has been presented in this paper

\section{Competing Interests}
We have no competing interests

\section{Authors' Contributions}
This author is the sole author of this paper

\section{Acknowledgements}
The author wishes to thank Prof. John R. Willis for his comments and suggestions.

\section{Funding}
The author acknowledges the support of the UCSD subaward UCSD/ONR W91CRB-10-1-0006 to the Illinois Institute of Technology (DARPA AFOSR Grant RDECOM W91CRB-10–1-0006 to
the University of California, San Diego). 


\end{document}